\documentclass[prb,twocolumn,amssymb,amsmath,aps,superscriptaddress,showpacs]{revtex4}
\usepackage{graphicx}
\usepackage{hyperref}
\usepackage{amsfonts}
\usepackage{epsfig}
\def\bea{\begin{eqnarray}}
\def\eea{\end{eqnarray}}
\def\be{\begin{equation}}
\def\ee{\end{equation}}

\begin{document}
\title{Vortex Penetration into a Type II Superconductor due to a Mesoscopic
External Current}
\author{Eran Sela and Ian Affleck \date{\today}}
\affiliation{Department of Physics and Astronomy, University of
British Columbia, Vancouver, B.C., Canada, V6T 1Z1}

\begin{abstract}
Applying the London theory we study curved vortices produced by an
external current near and parallel to the surface of a type II
superconductor. By minimizing the energy functional we find the
contour describing the hard core of the flux line, and predict the
threshold current for entrance of the first vortex. We assume that
the vortex entrance is allowed due to surface defects, despite the
Bean-Livingston barrier. Compared to the usual situation with a
homogeneous magnetic field, the main effect of the present geometry
is that larger magnetic fields can be applied locally before
vortices enter the superconducting sample. It is argued that this
effect can be further enhanced in anisotropic superconductors.
\end{abstract}

\maketitle
\section{Introduction}
\label{se:int} Surface barrier effects in type II superconductors
have been predicted by Bean and Livingston~\cite{Bean64} and de
Gennes.\cite{DeGennes66} The entry of flux lines into a planar type
II superconductor situated in an external magnetic field $H_{ext}$
parallel to its surface is opposed by a strong surface barrier when
$H_{ext} = H_{c1}$, the first critical field. Therefore the entry of
flux lines could occur at a field value $H_{ext} = H_{S} \sim H_{c2}
\gg H_{c1}$, where $ H_{c2}$ is the second critical field. These
surface barrier effects have been observed experimentally in the
$60$`s on lead thallium alloys~\cite{Joseph64} and on niobium
metal,\cite{Blois64} and make it difficult to measure directly
thermodynamic properties of the superconductor.

Typically surface barriers are reduced due to surface disorder,
which creates large local magnetic fields and allows for nucleation
of vortices. Suppression of surface barriers for flux penetration
was observed on YBaCuO~\cite{Yeshurun91} and in BiSrCaCuO
whiskers~\cite{Gregory01} due to heavy ion irradiation. In
ellipsoid-shaped YBaCuO it has been argued that due to roughness of
submicrometer order the surface barrier does not push the
penetration field $H_S$ above $H_{c1}$ but only lowers the rate of
vortex entry.\cite{Liang94}

Another source for the delay of the entry of flux lines into
superconductors is the ``geometrical
barrier",\cite{Zeldov94,Brandt99} which is particularly important in
thin films of constant thickness (i.e., rectangular cross section).
This effect is absent only when the superconductor is of exactly
ellipsoidal shape or is tapered like a wedge with a sharp edge where
flux penetration is facilitated. The resulting absence of hysteresis
in wedge-shaped samples was nicely shown by Morozov et
al.~\cite{Morozov97}

\begin{figure}[h]
\begin{center}
\includegraphics*[width=70mm]{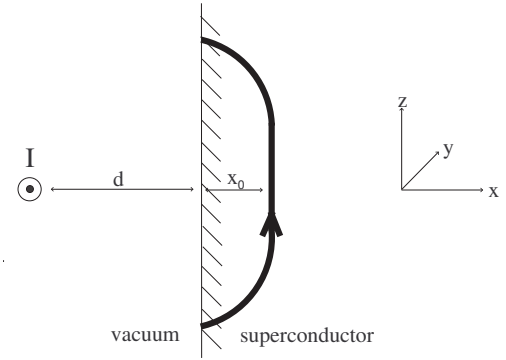}
\caption{Curved flux line near the surface of a superconductor
enabled by an external current $I$. \label{fg:1} }
\end{center}
\end{figure}
In this paper we study another source for delay of entrance of flux
lines, due to inhomogeneity of the external magnetic field. In
particular we consider magnetic field produced by an external
current $I$ flowing parallel to the surface of a type II
superconductor, see Fig. 1. The magnetic field produced by the
external current enters the sample as curved vortices at
sufficiently large current. We find that the entrance of the first
line occurs when the induced magnetic field at the surface at the
position closest to the wire already exceeds the bulk critical field
$H_{c1}$. This delay in entrance of the curved vortices occurs due
to geometrical reasons: The entry and outlet points are associated
with an energy cost $\sim \frac{\phi_0^2}{\mu_0 \lambda}$, where
$\lambda$ is the penetration depth, $\phi_0 $ is the flux quantum,
and $\mu_0$ is the free permeability. Note that $\phi_0^2/\mu_0 k_B
= 0.2464 ~10^6 K - \mu m$, implying that in typical superconductors
this is a large energy scale. In addition the spatially averaged
magnetic field experienced by the vortex is lower than the the
maximal one occurring closest to the wire. Considering those effects
in an actual calculation we find how large a magnetic field can be
applied locally without introducing vortices into the sample.

This implies that application of magnetic field by an external
current near the SC can be convenient for experiments demanding
sizable magnetic fields in the vortex-free state. As such an
experiment we mention the London-Hall effect.\cite{London50} Whereas
this effect was observed in regular
superconductors,\cite{Bok68,Brown68,Morris71} it is now interesting
to measure it in high temperature superconductors. Typically
$H_{c1}$ is quite low in these materials and therefore vortices
penetrate the sample at very low homogeneous magnetic fields; hence
our geometry can be useful. However other surface effects seem to be
an additional obstacle for the observation of the London effect in
high temperature superconductors.\cite{Lipavsky04}

A parameter which we leave out of consideration in this work is
anisotropy of the superconductor, which is particularly important in
high temperature layered superconductors. In the case of strong
anisotropy additional complications enter the problem even in the
case of uniform magnetic field, where the direction of the vortices
deviates from the direction of the external magnetic
field.\cite{Blatter93} For certain (elliptical) treatment of the
short distance cutoff the vortices can have two different
directions, corresponding to two degenerate minima in the free
energy.\cite{Sudbo93}

We argue that strong anisotropy is expected to have important
effects in our geometry, increasing further the maximal local
magnetic field allowed before curved flux lines penetrate the
sample. Consider the case where the $\hat{c}$ axis of a uniaxially
anisotropic superconductor corresponds to the direction $\hat{x}$ in
our geometry, $\hat{c}
\parallel \hat{x}$. In this case the surface of
the superconductor, parallel to the external wire, corresponds to an
ab-plane. In the limit of strong anisotropy $\lambda_{ab}\ll
\lambda_c$ the bulk critical field parallel to the surface $H_{c1}
\cong \frac{\phi_0}{4 \pi \mu_0 \lambda_{ab} \lambda_c} \log
\frac{\lambda_c}{\xi}$ becomes very small. On the other hand, the
entry and outlet points of the flux line are associated with a large
energy cost $\sim \frac{\phi_0^2}{\mu_0 \lambda_{ab}}$ independent
of $\lambda_c$. Therefore we expect the maximal surface magnetic
field before the entry of the first vortex to increase relative to
$H_{c1}$ as a function of $\lambda_c/\lambda_{ab}$. We leave a
detailed treatment of anisotropy in this geometry for a future work.

The paper is organized as follows. In Sec.~\ref{se:Formulation} we
formulate the problem and obtain expressions for the magnetic field
and free energy within London theory. In Sec.~\ref{se:Barrier} we
present and discuss the numerical results for the minimization of
the free energy as a function of vortex contour. In
Sec.~\ref{se:Formulation} and Sec.~\ref{se:BarrierA} we consider the
simpler but unrealistic case of a wire with zero width (i.e. $\ll
\lambda$). In Sec.~\ref{se:BarrierB} we generalize to wires with
finite width. Sec.~\ref{se:Conclusions} contains conclusions. Some
details about the derivation of the expression of the free energy
are relegated to the appendix.

\section{Formulation}
\label{se:Formulation} 
\begin{figure}[h]
\begin{center}
\includegraphics*[width=45mm]{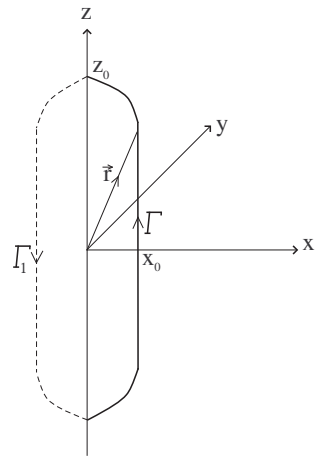}
\caption{$\Gamma$ is the axial line of the vortex line. The closed
contour $\gamma$ is $\Gamma+\Gamma_1$. \label{fg:axialline} }
\end{center}
\end{figure}

Suppose that a type-II superconductor (SC)
occupies the region $x>0$ and magnetic field is induced by an
external current $I$ flowing along a wire of zero cross-section at
$(x,z)=(-d,0)$, see Fig.~(\ref{fg:1}). Our main object under
consideration is a curved flux line lying in the plane $y=0$. Let
$\gamma$ denote the closed contour in Fig.~(\ref{fg:axialline})
consisting of the axial line of the flux line $\Gamma$ and a line
$\Gamma_1$ symmetric to $\Gamma$ with respect to the plane $x=0$,
corresponding to an image vortex. Upon further increasing the
current a lattice of curved vortices is expected to form along the
wire. However here we shall concentrate on small currents and a
single flux line.

In the type II limit, where the coherence length $\xi$ is much
shorter than the penetration depth $\lambda$, the total free energy
at zero temperature is given by~\cite{DeGennes66}
\begin{eqnarray}
\label{eq:F} F[\Gamma]&=&\frac{\mu_0}{2} \int_{r>\xi} d^3 r
\bigl[\vec{H}^2+\theta(x) \lambda^2 (\vec{\nabla} \times \vec{H})^2
\bigr] \nonumber \\
&-&\mu_0 \int d^3 r  \vec{A} \cdot \vec{j}_{\rm{ext}}.
\end{eqnarray}
Here $\vec{j}_{\rm{ext}}=-I \delta(x+d) \delta(z) \hat{y}$, $I$ is
the applied current through the wire, $\vec{A}$ is the vector
potential $\vec{H} = \vec{\nabla} \times \vec{A}$ and $\theta(x)$ is
the unit step function. The integral $\int_{r>\xi}$ is carried out
in all space outside of the vortex ``hard core" $\Gamma$. We assume
that the radius of curvature of $\Gamma$ is larger than $\xi$ at any
point in $\Gamma$. Note that at $x=0$ there is an apparent kink in
$\gamma$, however this should be though of as a kink only for length
scales large compared to $\xi$.

The corresponding equations for the magnetic field $\vec{H}$ are the
Maxwell equation, $\vec{\nabla} \times \vec{H}=\vec{j}_{\rm{ext}}$
for $x<0$, and the London equation, $(1-\lambda^2
\vec{\nabla}^2)\vec{H}(\vec{r}) = \frac{\phi_0}{\mu_0} \int_{\Gamma}
d\vec{r}' \delta^3(\vec{r}-\vec{r}') $ for $x>0$. For all $x$ we
also have $\vec{\nabla} \cdot \vec{H}=0$. In addition we impose
appropriate boundary conditions at $x=0$: the magnetic field is
continuous, and no supercurrent flows perpendicular to the surface:
$\vec{j}_x = (\vec{\nabla} \times \vec{H})_x=0$. To construct a
solution we use the functions
\begin{eqnarray}
\label{eq:functions}
\vec{H}^{hom}_{A(\vec{k}_2),B(\vec{k}_2)}(\vec{r})&=&\int \frac{d^2
k_2}{(2 \pi)^2}e^{i \vec{k}_2 \cdot \vec{r}}\times \nonumber \\
&\times& \left \{
\begin{array}{ll}
A(\vec{k}_2)[-k_2^2 \hat{x}+i \vec{k}_2 \tau(k_2)]e^{-\tau(k_2) x} & x> 0  \\
B(\vec{k}_2)[k_2 \tau(k_2)  \hat{x}+i \vec{k}_2 \tau(k_2)]e^{k_2 x}
& x< 0
\end{array} \right. , \nonumber \\
\vec{H}_{\gamma}(\vec{r})&=&\frac{\phi_0}{\mu_0}\int_{\gamma}
d\vec{r}' \int \frac{d^3k}{(2 \pi)^3} e^{i \vec{k}\cdot
(\vec{r}-\vec{r}')}
\frac{1}{1+\lambda^2 k^2}, \nonumber \\
\vec{H}_{I',d'}(\vec{r})& =&\frac{I'}{2 \pi} \frac{(-z,0,x +
d')}{(x+ d')^2+z^2}.
\end{eqnarray}
Here $\vec{k}_2=k_y \hat{y}+k_z \hat{z}$, $k_2 =
\sqrt{k_y^2+k_z^2}$, and $\tau(k)=\sqrt{k^2+\lambda^{-2}}$. For any
$A(\vec{k}_2),B(\vec{k}_2)$, the function $\vec{H}^{hom}$ satisfies
the homogeneous equations $\vec{\nabla} \times \vec{H}^{hom}=0$ for
$x<0$, and $(1-\lambda^2 \vec{\nabla}^2)\vec{H}^{hom} =0 $ for
$x>0$. The function $\vec{H}_{\gamma}$ satisfies the London equation
$(1-\lambda^2 \vec{\nabla}^2)\vec{H}_{\gamma}(\vec{r}) =
\frac{\phi_0}{\mu_0} \int_{\gamma} d\vec{r}'
\delta^3(\vec{r}-\vec{r}') $ in all space. The function
$\vec{H}_{I',d'}$ satisfies Maxwell equation $\vec{\nabla} \times
\vec{H}_{I',d'}(\vec{r})=\vec{j}'_{\rm{ext}}$ for
$\vec{j}'_{\rm{ext}}=I' \delta(x+d') \delta(z) \hat{y}$ for all
space.

\begin{figure}[h]
\begin{center}
\includegraphics*[width=40mm]{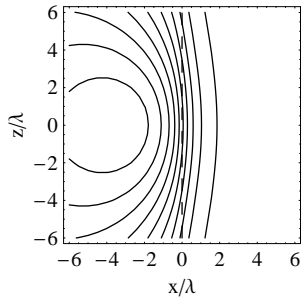}
\caption{Field lines of the vortex-free solution $\vec{H}_0(x,z)$
for $d=5 \lambda$ [direction of field lines correspond to
anticlockwise rotation around $(x,z)=(-d,0)$].\label{fg:1a} }
\end{center}
\end{figure}

Defining the surface 2-dimensional Fourier transform
$\vec{H}_\mu^{surf}(\vec{k}_2) =  \int dy dz e^{-i \vec{k}_2 \cdot
\vec{r}} \vec{H}_\mu (0,y,z)$, for $\mu = \gamma, \{I',d' \}$, one
finds
\begin{eqnarray}
\vec{H}_\gamma^{surf}(\vec{k}_2) = \frac{\phi_0}{2\mu_0 \lambda^2}
\int_\gamma d\vec{r}' e^{-i \vec{k}_2 \cdot \vec{r}'}
\frac{e^{-\tau(k_2) |r_x'|}}{\tau(k_2)}, \nonumber \\
\vec{H}_{I,d}^{surf}(\vec{k}_2) =\delta(k_y)\pi I e^{-|k_z d|}(i~
{\rm{sgn}}~k_z,0,{\rm{sgn}}~d).
\end{eqnarray}
The solution of the equations satisfying the desired boundary
conditions is obtained by adding together the functions in
Eq.~(\ref{eq:functions}), and solving for $A(\vec{k}_2)$ and
$B(\vec{k}_2)$ to give continuity. It is convenient to include an
image current at $x=-d$. The total magnetic field is
\begin{eqnarray}
\label{eq:H}
\vec{H}&=&\vec{H}_0+\vec{H}_v+\vec{H}_s ,\nonumber \\
\vec{H}_0
&=&\theta(-x)(\vec{H}_{I,d}+\vec{H}_{-I,-d})+\vec{H}_{s0},~~~\vec{H}_{s0}=\vec{H}^{hom}_{A_0,B_0},\nonumber \\
\vec{H}_v &=&\theta(x)\vec{H}_{\gamma},~~~~ \vec{H}_s
=\vec{H}^{hom}_{A_1,B_1},
\end{eqnarray}
where
\begin{eqnarray}
A_0(\vec{k}_2)&=&\frac{2[\vec{H}_{I,d}^{surf}(\vec{k}_2)]_z}{i
k_z(\tau(k_2)+k_2)},
~~~B_0(\vec{k}_2)=-A_0(\vec{k}_2)\frac{k_2}{\tau(k_2)},\nonumber \\
A_1(\vec{k}_2)&=&B_1(\vec{k}_2)=\frac{[\vec{H}_{\gamma}^{surf}(\vec{k}_2)]_x}{k_2[k_2+\tau(k_2)]}.
\end{eqnarray}
In the absence of vortices the magnetic field is given by
$\vec{H}_0$. It is plotted in Fig.~(\ref{fg:1a}) for $d=5 \lambda$.

The total free energy as function of $\Gamma$ is obtained by
substituting the magnetic field Eq.~(\ref{eq:H}) into the free
energy Eq.~(\ref{eq:F}). We obtain
\begin{equation}
\label{eq:Fterms} F =F_0+ F_v + F_{s} + F_{ext}.
\end{equation}
Here
\begin{eqnarray}
\label{eq:Ftot} F_0&=&\frac{\mu_0}{2} \int d^3 r
\bigl[\vec{H}_{0}^2+\theta(x) \lambda^2(\vec{\nabla} \times
\vec{H}_{0})^2 -2\vec{A}_0 \cdot \vec{j}_{\rm{ext}}\bigr], \nonumber \\
F_{i}&=&\frac{\mu_0}{2} \int d^3 r \bigl[\vec{H}_{i}^2+\theta(x)
\lambda^2 (\vec{\nabla} \times \vec{H}_{i})^2 \bigr],~~~i=v,s,
\nonumber
\\F_{ext}&=&-\mu_0 \int d^3 r  (\vec{A}_v + \vec{A}_{s} )\cdot
\vec{j}_{\rm{ext}} .
\end{eqnarray}
Here $\vec{H}_i=\vec{\nabla} \times \vec{A}_i$, $i=0,v,s$. All mixed
terms between $\vec{H}_0,\vec{H}_v,\vec{H}_{s}$ vanish. For the
vanishing of mixed terms involving $\vec{H}_v$ see
Ref.~\onlinecite{Brandt80}, p.579. We prove the vanishing of the
remaining crossed terms between $\vec{H}_0$ and $\vec{H}_s$ in the
appendix.

\begin{figure}[h]
\begin{center}
\includegraphics*[width=75mm]{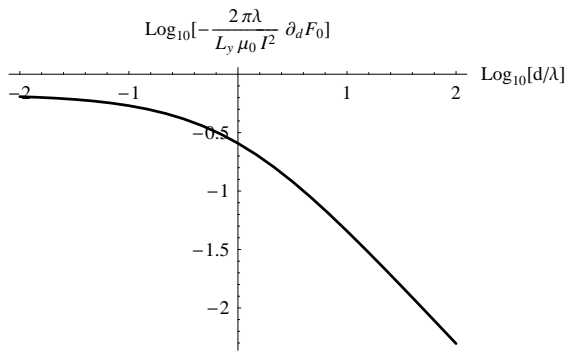}
\caption{Repulsive force between the superconductor and the wire in
the absence of vortices. The dimensionless force $-(2 \pi
\lambda/L_y \mu_0 I^2)
\partial_d F_0$ behaves as $\lambda/(2d)$ for $d \gg \lambda$, and goes to a constant $c_0 \sim
0.665$ for $d \ll \lambda$.\label{fg:Force} }
\end{center}
\end{figure}

The term $F_0$ is the energy of the system without vortices. To
evaluate it we introduce a finite wire radius $a \ll \lambda, d$,
and assume the external current flows in a thin shell of this
radius. Note that $F_0$ scales linearly with the length of the wire,
$L_y$. The result of a calculation, using the methods of the
appendix, is
\begin{eqnarray} \frac{F_0}{L_y}&=&-\frac{\mu_0 I^2}{2 \pi} \left[
\frac{1}{2} \log(2d/a)+g(d/\lambda) \right], \nonumber
\\
g(y)&=&\int_0^\infty dx \frac{e^{-2 x}}{x+\sqrt{x^2+y^2}}.
\end{eqnarray}
We can infer from it the repulsive force per unit length $\frac{
\partial_d F_0}{L_y}<0$ between the wire and the SC. It is plotted in Fig.~(\ref{fg:Force}) (for
$a/\lambda = 0.01$). Using $g(y \rightarrow \infty)  =\frac{1}{2
y}$, $g(y \rightarrow 0)= \frac{1}{2 }\log\frac{1}{2 y}$, we may
identify two regimes. (i) $d \gg \lambda$: Here  $g \to 0$ and $
\frac{
\partial_d F_0}{L_y} \rightarrow - \frac{\mu_0 I^2}{2 \pi (2 d)}$. In agrement with Ampere force law, this corresponds to a repulsive
force per unit length between two wires $2d $ apart carrying current
$I$ with opposite direction. This is the origin of the levitation
effect. The second wire corresponds to the term $\vec{H}_{{-I,-d}}$
in the solution for $\vec{H}_0$, see Eq.~(\ref{eq:H}); (ii) $d \ll
\lambda$: The $1/d$ divergence in the force is cutoff by $\lambda$.
The limiting repulsion force per unit length as the wire approaches
the surface is $\frac{
\partial_d F_0}{L_y} \rightarrow - \frac{\mu_0 I^2}{2 \pi \lambda} c_0$ where $c_0 \sim
0.665$.

The term $F_{ext}$ accounts for the interaction between the vortex
and the external current. Using $\vec{j}_{\rm{ext}}=-I \delta(x+d)
\delta(z) \hat{y}$ we have
\begin{eqnarray}
F_{ext}=\mu_0 I \int_{-\infty}^\infty d y  [\vec{A}_v(-d,y,0) +
\vec{A}_{s} (-d,y,0)]_y .
\end{eqnarray}
The contour of integration $(x,y,z)=(0,-\infty,0)$ $\rightarrow
(0,\infty,0)$ corresponds to the external current. Physically the
wire should be closed into a loop, and we may close the contour of
integration e.g. in the $xy$ plane from the $x \rightarrow -\infty$
side. Then, using Green's theorem we obtain
\begin{eqnarray}
F_{ext}=\mu_0 I \int_{-\infty}^{-d} dx \int_{-\infty}^\infty d y
 [\vec{H}_v(x,y,0) + \vec{H}_{s} (x,y,0)]_z .\nonumber
\end{eqnarray}
Note that $\vec{H}_v(x,y,0)$ vanishes at $x<0$. Using the formula
for $\vec{H}_{s}$, Eq.~(\ref{eq:H}), we obtain
\begin{eqnarray}
\label{eq:FI} F_{ext}&=&- \frac{I \phi_0}{\pi} \int_{\Gamma}
d\vec{r}_z \int_0^\infty dk e^{- k d } \cos(k r_z) \times
\nonumber \\
&\times& \left( 1-e^{- \tau(k) r_x}\right) \left(
1-\frac{k}{\tau(k)}\right).
\end{eqnarray}
We used the identity $\oint_\gamma d \vec{r} \cdot \vec{\nabla}
\mathcal{F}(\vec{r})=0$ which holds for any continuous function
$\mathcal{F}$ if $\gamma$ is a closed contour. In this calculation
$\mathcal{F}(\vec{r}) = e^{i k r_z}{\rm{sgn}}(r_x) \left( 1-e^{-
\tau(k) |r_x|}\right)$.

The terms $F_v$ and $F_s$ have been derived in
Refs.~[\onlinecite{Brandt80,Shehata84}],
\begin{eqnarray}
\label{eq:Fvs} F_v&=&\frac{\phi_0^2}{2 \mu_0} \sum_{i=x,y,z}
\int_\gamma d\vec{r}_i \int_\gamma d\vec{r}'_i
\frac{\exp(-|\vec{r}-\vec{r}'|/
\lambda)}{8 \pi \lambda^2 |\vec{r}-\vec{r}'|}  , \nonumber \\
F_s&=&\frac{\phi_0^2}{2 \mu_0} \int_\Gamma d\vec{r}_z
\int_{\Gamma_1} d\vec{r}'_z V^{(s)}(\vec{r} - \vec{r}').
\end{eqnarray}
The term $F_v$ is sensitive to the short distance cutoff $\xi$. To
account for the cutoff we restrict the contour integration to
$|\vec{r} - \vec{r}'|>\xi$. Here the anisotropic kernel for $F_s$ is
\begin{eqnarray}
V^{(s)} (\vec{r})= \frac{1}{2 \pi \lambda^2 } \int_0^\infty dk
\left( 1 - \frac{k}{\tau(k)}\right)e^{- \tau(k) |r_x|} J_0(k
|r_z|),\nonumber
\end{eqnarray}
where $J_0(x)$ is a Bessel function, and this integral can be done
and expressed in terms of other Bessel functions.\cite{Shehata84}
Note that $V^{(s)}(r_x \to 0,r_z \to 0)= (2 \pi \lambda^3)^{-1}$,
hence there is no need to regulate $F_s$ with a cutoff.

\begin{figure}[h]
\begin{center}
\includegraphics*[width=30mm]{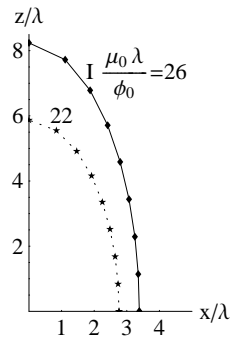}
\caption{Contours corresponding to local minimum of $F$ for $d=10
\lambda$, and for the specified currents. We assume that $\Gamma$
has the reflection symmetry $z\rightarrow -z$, and plot $\Gamma$
only for $z \ge 0$. \label{fg:contour} }
\end{center}
\end{figure}

Different than the usual case with a uniform magnetic field, in our
problem the energy $F =F_0+ F_v + F_{s} + F_{ext}$ is a function of
the contour $\Gamma$ and is minimized for a particular contour which
we need to find. To this end we minimize $F[\Gamma]$ numerically,
approximating $\Gamma$ by a polyline having $2M$ equal length sides
($M=8$ in most simulations). We assume that $\Gamma$ has the
reflection symmetry $z\rightarrow -z$. This leads to a
$M+1$-dimensional parameter space in which we search for the minimum
of $F$. For an example see Fig.~(\ref{fg:contour}). In all our
calculations $\xi=.001 \lambda$.

\section{Surface barrier}
\label{se:Barrier} We find that the free energy contains a surface
energy barrier. From this section we shall disregard the vortex
independent term of the free energy, $F \to F_v + F_{s} + F_{ext}$.
For later comparison we briefly discuss the case with homogeneous
magnetic field.\cite{Bean64} Consider a semi-infinite type-II
superconductor with a flux thread within it, parallel to the surface
and to the external magnetic field $H_{ext}$ ($\parallel \hat{z}$).
The line energy $f=F/L_z$ ($L_z$ is the length of the vortex taken
to be parallel to $\hat{z}$) as function of the distance from the
surface $x_0$, is given by~\cite{Bean64,DeGennes66}
\begin{equation}
\label{degenes} f(x_0)=\phi_0 \bigl[ H_{ext}
e^{-x_0/\lambda}-\frac{1}{2}h(2 x_0)+H_{c1}-H_{ext} \bigr].
\end{equation}
Here $h(r) = \frac{\phi_0}{2 \pi \mu_0 \lambda^2} K_0\bigl(
\frac{r}{\lambda} \bigr)$ is the function giving the field at
distance $r$ of a single straight flux line,
$H_{c1}=\frac{1}{2}h(\xi) \cong \frac{\phi_0}{4 \pi \mu_0 \lambda^2}
\log \frac{\lambda}{\xi}$, and $K_0$ is the zero-order Bessel
function. The term $\phi_0 H_{ext} e^{-x_0/\lambda}$ describes the
interaction of the line with the external field and associated
screening currents. It is a repulsive term. The term $-\phi_0 h(2
x_0) / 2$ represents the attraction between the line and its image.
When $H_0 \sim H_{c1}$ there is a strong barrier opposing the entry
of a line. We can understand this barrier as follows: When
$H_{ext}=H_{c1}$, $f(x_0=0)=f(x_0 = \infty)=0$. If we start from
$x_0$ large and bring the line closer to the surface, the repulsive
term ($\sim e^{-x_0/\lambda}$) dominates the image term ($\sim e^{-2
x_0/\lambda}$). Thus $f$ becomes positive and we have a barrier. The
barrier disappears, however, in high fields. When $H
> H_S = \phi_0/4 \pi \lambda \xi$, the slope $\partial f/\partial x_0|_{x_0 \sim \xi}$ becomes
negative. $H_S$ is of the order of the thermodynamic critical field
$H_{c2}$. The conclusion is that, at field $H < H_S$, the lines
cannot enter in an ideal specimen (although their entry is
thermodynamically allowed as soon as $H>H_{c1}$). However this
picture is modified in experiment due to surface inhomogeneities
producing local large magnetic fields, and allowing vortices to
enter the sample above $H_{c1}$.

\subsection{Results for wire with zero width}
\label{se:BarrierA} 
\begin{figure}[h]
\begin{center}
\includegraphics*[width=75mm]{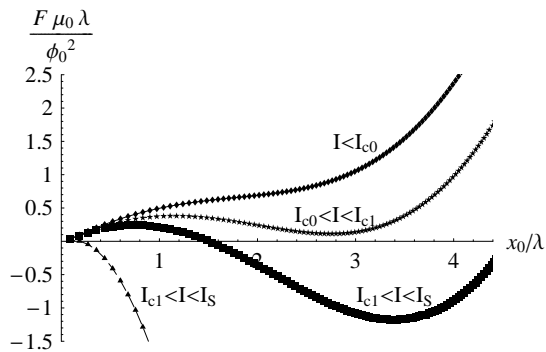}
\caption{Evolution of surface barrier as function of external
current for $d/\lambda=10$. When $I<I_{c0}$ (diamonds, $I=19
\frac{\phi_0}{\mu_0 \lambda}$) the force on the line points always
towards the surface. When $I_{c0}<I<I_{c1}$ (stars, $I=22
\frac{\phi_0}{\mu_0 \lambda}$) there exists a meta-stable minima
with positive energy. When $I_{c1}<I<I_{S}$ (squares and triangles
$I=26,80 \times \frac{\phi_0}{\mu_0 \lambda}$) the minimum energy is
negative, but a barrier opposes the entry of the flux line. In each
point in this plot we have minimized numerically $F$ with respect to
the contour $\Gamma$ at fixed $x_0$. \label{fg:EL}}.
\end{center}
\end{figure}

We find a similar energy barrier for the
entrance or exit of a curved vortex in our geometry with an external
current rather than an homogeneous external magnetic field.  This
barrier can be visualized in the curves in Fig.~(\ref{fg:EL})
(except for the diamonds). Note that typically the barrier hight
$\Delta$ is of order $\Delta \sim \phi_0^2/\mu_0 \lambda \gg T_c$,
where $T_c$ is the critical temperature of the SC. This implies
rather small tunneling probabilities $e^{-\Delta/T} \ll 1$ which
prevents entry of vortices for clean surfaces. However for strong
disorder, vortices can enter more efficiently via nucleation at
impurity sites. The contours corresponding to the minimum of the
curves with stars and squares are plotted in
Fig.~(\ref{fg:contour}). In all our calculations $\xi=.001 \lambda$.

Figure (\ref{fg:EL}) implies the following picture. For
infinitesimal current there is no stable vortex configuration. As
the current increases we identify three threshold currents
$I_{c0}<I_{c1}<I_{S}$: When the current exceeds $I_{c0}$ a
meta-stable minima with $F>0$ occurs. When the current exceeds
$I_{c1}$, the minimum energy changes sign, $F<0$, but still there is
an energy barrier for the entrance of a flux line. When the current
exceeds $I_{S}$ the barrier disappears.

\begin{figure}[h]
\begin{center}
\includegraphics*[width=80mm]{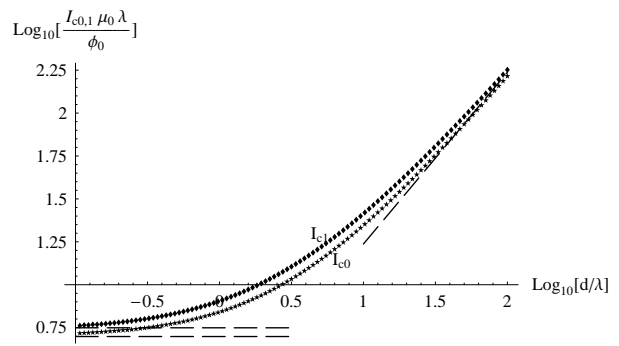}
\caption{Dependence of threshold currents $I_{c0}$ and $I_{c1}$ on
$d/\lambda$. \label{fg:Ic01}}.
\end{center}
\end{figure}

In Fig.~(\ref{fg:Ic01}) we investigated the dependence of $I_{c0}$
and $I_{c1}$ on the distance to the wire $d$. In the limit $d \gg
 \lambda $ the results for $I_{c1}$ are consistent with the formula $I_{c1} \to \pi d H_{c1}$ [see diagonal dashed line in
Fig.~(\ref{fg:Ic01})]. The behavior of $I_{c0}$ in that limit shows
that the region of metastability $I_{c0}<I<I_{c1}$ is very narrow.
This behavior appears in sharp contrast to the case of uniform
magnetic field even in the limit $d \gg
 \lambda$: We recall that Eq.~(\ref{degenes}) predicts metastable solutions
for infinitesimal homogeneous magnetic field $H_{ext}$. These states
live far from the surface as $H_{ext}$ becomes smaller. This effect
is absent in our geometry both due to the fact that the effective
external magnetic field created by the wire decays at long distances
from the surface and due to the line energy for penetration a long
distance into the SC. In the other extreme limit $d \ll \lambda$ we
observed from the numerical solution that the contour $\gamma$ can
be approximated by a circle centered at the origin. Making this
assumption we can calculate $I^{circle}_{c0}=10^{0.6981}
\frac{\phi_0}{\mu_0 \lambda}$ ($x_0 \sim 0.72 \lambda$, $F \mu_0
\lambda/\phi_0^2=0.1715$), and $I^{circle}_{c1}=10^{0.749}
\frac{\phi_0}{\mu_0 \lambda}$ ($x_0=1.27 \lambda$, $F=0$) in the
limit $d \rightarrow 0$. This approximation is in reasonable
agreement with the actual solution as the horizontal dashed lines
show.

\begin{figure}[h]
\begin{center}
\includegraphics*[width=90mm]{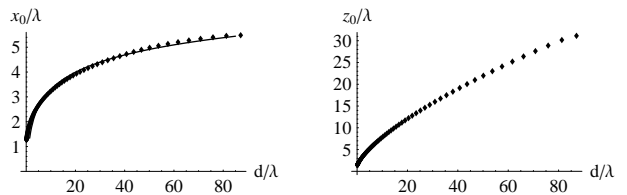}
\caption{Extensions of the curved flux line along $x$ and $z$ as
function of $d/\lambda$, at $I = I_{c1}$. \label{fg:xz}}.
\end{center}
\end{figure}

The shape of the contour changes as function of $d$. In
Fig.~(\ref{fg:xz}) we plot the extension of the contour in the $x$
and $z$ directions. We fitted the numerical results for $x_0$ with
an empirical formula $x_0/\lambda = c +\log(d /\lambda)$ with $c
\sim 1$, implying that the penetration of the vortex is of order
$\lambda$ for all $d$. On the other hand it appears that $z_0$ grows
linearly as function of $d$. In the limit $d \to 0$ we have
$x_0/\lambda \to 1.26$  and $z_0/\lambda \to 1.43$ (implying that
the circular contour is only an approximation).

\begin{figure}[h]
\begin{center}
\includegraphics*[width=80mm]{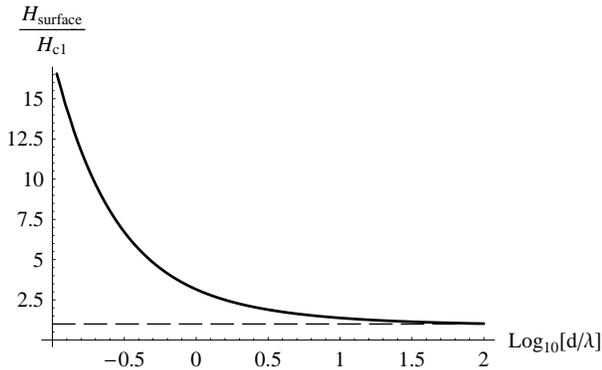}
\caption{Magnetic field at the surface (see definition in the text)
of a clean SC just before the entry of the first vortex for current
approaching $I_{c1}$. At $d \gg \lambda$ this magnetic field tends
to the bulk critical field $H_{c1}$. As $d$ becomes smaller the SC
can sustain larger magnetic fields in the vortex-free (Meissner)
state.\label{fg:Hratio}}
\end{center}
\end{figure}

For disordered surfaces, the present geometry can be useful for
application of large magnetic fields on a SC sample in a vortex free
state. The maximal magnetic field that can be applied in a vortex
free state using the wire geometry is $\vec{H}_0[x=z=0]$ [see
Eq.~(\ref{eq:H})] at current $I_{c1}$. In the limit $d \gg \lambda$
this magnetic field coincides with the bulk first critical field
$H_{c1} \approx \frac{\phi_0}{4\pi \mu_0
\lambda^2}\log(\lambda/\xi)$, however at smaller $d$ the magnetic
field at the surface increases. This is shown in
Fig.~(\ref{fg:Hratio}) where we plot the magnetic field
$H_{surface}=[\vec{H}_0(0,0,0)]_z = [\vec{H}_{s0}(0^+,0,0)]_z$ given
in Eq.~(\ref{eq:H}) at the current $I_{c1}$, which we calculated
above as function of $d$. Note that the field enhancement is small
for $d>3\lambda$ ($H_{surface} \cong 2 H_{c1}$ for $d=3 \lambda$).

\begin{figure}[h]
\begin{center}
\includegraphics*[width=70mm]{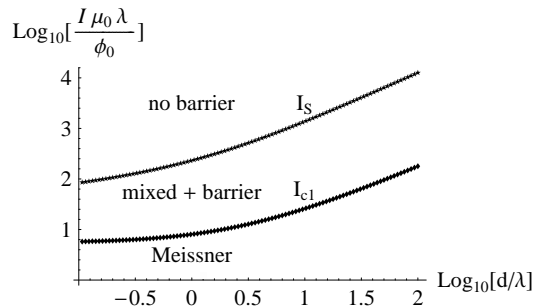}
\caption{Phase diagram. For $I<I_{c1}$ vortices are
thermodynamically unfavorable. For $I>I_{c1}$ curved flux lines
become favorable, but an energy barrier opposes their entry until
the current exceeds $I_S$. A metastable phase exists in a narrow
strip below the boundary $I=I_{c1}$, shown in Fig.~(\ref{fg:Ic01}).
\label{fg:Ic2}}.
\end{center}
\end{figure}

We turn to an estimation of the threshold current $I_S$ at which the
barrier disappears. A more precise calculation would involve the
Ginzbur-Landau theory. We follow the above analysis of
$H_S$.~\cite{DeGennes66} Since the London theory is applicable at
distances $\gg \xi$ we estimate $I_S$ using
\begin{equation}
\label{eq:zeroderivative} \frac{\partial F }{\partial x_0}
\bigl|_{x_0 \sim \xi}=0.
\end{equation}
We find numerically that at $x_0 \ll \lambda$ the closed contour
$\gamma$ is well approximated by a circle with radius $x_0$ centered
at $x=z=0$. In the limit $x_0 \sim \xi \ll \lambda$ we can evaluate
the functional $F$ analytically as function of $x_0$. In
Eq.~(\ref{eq:Fvs}) for $F_v$ we can set $\exp(-|\vec{r}-\vec{r}'|/
\lambda) \to 1$, hence
\begin{eqnarray}
F_v(x_0) &\sim&  \frac{\phi_0^2 x_0}{32 \pi \mu_0 \lambda^2}
\int_0^{2\pi} d \theta_1 \int_0^{2\pi} d \theta_2
\frac{\cos(\theta_1 -\theta_2)}{\left| \sin  \frac{\theta_1
-\theta_2}{2}\right|} \times
\nonumber \\
&\times& \theta(2 x_0 \left| \sin  \frac{\theta_1
-\theta_2}{2}\right|-\xi).
\end{eqnarray}
Compared to $F_v$, the stray term is negligibly small, $F_s \sim
\frac{\phi_0^2}{\mu_0 \lambda}\left( \frac{x_0}{\lambda} \right)^2$.
The interaction energy with the external current reads
\begin{eqnarray}
F_{ext}(x_0)&=& - \frac{I \phi_0 x_0^2}{2 \lambda^2}
\tilde{f}(d/\lambda),\nonumber \\
\tilde{f}(x)&=&\int_0^\infty dy e^{-y x} (\sqrt{y^2+1}-y).
\end{eqnarray}
Using these formulas for $F_{v}$ and $F_{ext}$ we obtain from
Eq.~(\ref{eq:zeroderivative}) the estimate
\begin{equation}
I_{S} \sim \frac{\phi_0}{8 \mu_0 \xi \tilde{f}(d/\lambda)}.
\end{equation}
The dependence of $I_S$ on $d/\lambda$ is hidden in the function
$\tilde{f}(x)$, with $\tilde{f}(x\to \infty)\to x^{-1}$,
$\tilde{f}(x\to 0)\to \log( x^{-1/2})+c$, where $c \sim 0.3$. For
$d\gg \lambda$ we have $I_{S} \sim \frac{\phi_0 d}{8 \mu_0 \xi
\lambda} $. In this case the magnetic field due to the external
current at $x=z=0$ is $[H_0(\vec{r}=0)]_z \to \frac{I_{S}}{\pi d}$.
It is of the order of the second critical field $H_{c2}$. In the
other limit $d\ll \lambda$ we have $I_{S} \sim \frac{\phi_0 }{4
\mu_0 \xi \log[\lambda/d]}$. Note that this behavior holds for $ \xi
\ll d \ll \lambda$. In this regime we have $I_S \gg I_{c1} \sim
\frac{\phi_0 }{ \mu_0 \lambda }$. In Fig.~(\ref{fg:Ic2}) we plot the
phase diagram.

\subsection{Finite wire cross section}
\label{se:BarrierB} 
\begin{figure}[h]
\begin{center}
\includegraphics*[width=60mm]{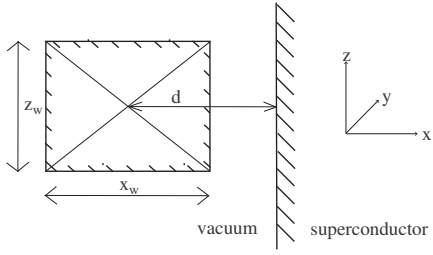}
\caption{Rectangular wire cross section. \label{fg:rec}}.
\end{center}
\end{figure}

\begin{figure}[h]
\begin{center}
\includegraphics*[width=90mm]{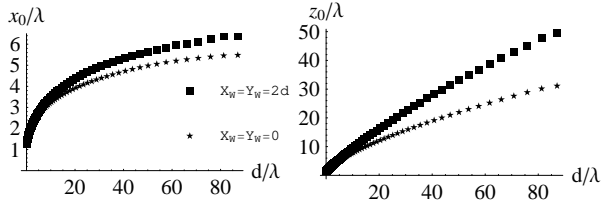}
\caption{Dependance on $d$ of the lengths $x_0$ and $z_0$,
characterizing the vortex contour, for zero versus finite wire cross
section [$x_x=z_w=0$ (stars) and $x_x=z_w=2d$ (squares)].
\label{fg:xzrec}}.
\end{center}
\end{figure}

\begin{figure}[h]
\begin{center}
\includegraphics*[width=70mm]{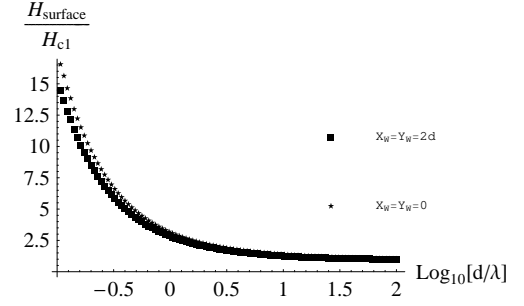}
\caption{Magnetic field at the surface (see definition in the text)
just before the entry of the first vortex at $I \to I_{c1}$, for
either zero wire cross section ($x_x=z_w=0$, stars) or finite cross
section ($x_x=z_w=2d$, squares). \label{fg:HratioREC}}
\end{center}
\end{figure}

Experimentally the wire carrying the external
current has a finite cross section. Therefore it is important to
include this effect in our calculations. Consider the rectangular
cross section as shown in Fig.~(\ref{fg:rec}), and assume current
$I$ flows uniformly in this cross section.
We can write the external current as
\begin{equation}
\label{eq:jext} \vec{j}_{\rm{ext}} \to -\frac{I}{x_w z_w}
\int_{d-\frac{x_w}{2}}^{d+\frac{x_w}{2}} d \tilde{d}
\int_{-\frac{z_w}{2}}^{\frac{z_w}{2}} d \tilde{z}
\delta(x-\tilde{d}) \delta(z-\tilde{z}) \hat{y}.
\end{equation}
The modification to the magnetic field
$\vec{H}=\vec{H}_0+\vec{H}_v+\vec{H}_s$ occurs only in the first
term,
\begin{equation}
\vec{H}_0(\vec{r}) \to \frac{1}{x_w z_w}
\int_{d-\frac{x_w}{2}}^{d+\frac{x_w}{2}} d \tilde{d}
\int_{-\frac{z_w}{2}}^{\frac{z_w}{2}} d \tilde{z}
\left[\vec{H}_0(x,y,z-\tilde{z}) \right]_{d \to \tilde{d}},\nonumber
\end{equation}
where $\vec{H}_0(x,y,z)$ is given in Eq.~(\ref{eq:H}). Next we focus
on the modification of the vortex dependent part of the free energy
$F = F_v + F_{s} + F_{ext}$. Only the term $F_{ext}$ is modified.
Using Eq.~(\ref{eq:jext}) it is easy to find that
\begin{eqnarray}
\label{eq:Fext} F_{ext} &\to& - \frac{I \phi_0}{\pi} \int_{\Gamma} d
\vec{r}_z \int_0^\infty dk e^{- k d } \cos(k r_z) (1-e^{- \tau(k)
r_x})\times \nonumber \\
&\times&\left( 1-\frac{k}{\tau(k)} \right) \left( \frac{\sinh\frac{k
x_w}{2}}{\frac{k x_w}{2}} \frac{\sin\frac{k z_w}{2}}{\frac{k
z_w}{2}} \right).
\end{eqnarray}
Let us first specialize to the case of square cross section where
the wire touches the SC, $x_w=z_w=2d$, and compare this with a point
like cross section $x_w=z_w \to 0$ which we considered until now (We
ignore any electron or Cooper pair tunneling between the SC and
wire). First we repeated the calculation of $I_{c0}$ and $I_{c1}$.
The results are roughly the same for both cross sections for $d
\lesssim \lambda$, and deviations up to $10 \%$ are obtained for $d
\gg \lambda$ up to $d=100 \lambda$. In Fig.~(\ref{fg:xzrec}) we
compare the contours at $I_{c1}$ as function of $d$ for the two
cross sections. We can see that $z_0$ changes by a factor of up to
$1.6$ for $d \le 90 \lambda$.
The magnetic field at the surface just below $I_{c1}$ is compared
for the two cross sections in Fig.~(\ref{fg:HratioREC}). At $d\gg
\lambda$ it approaches $H_{c1}$ in both cases, while for $d \ll
\lambda$ it is larger for point like cross section by about $10 \%$.

We estimate the typical values of the threshold current $I_{c1}$.
For the regime of interest $d \sim \lambda$, we have $I_{c1}$ of the
order of $\frac{\phi_0}{\mu_0 \lambda} = \frac{1.6455
mA}{\lambda[\mu m]}$. For $\lambda = 1 \mu m$ this corresponds to
current density of $\sim 1 mA  ~-~ \mu m^{-2}$.

\begin{figure}[h]
\begin{center}
\includegraphics*[width=60mm]{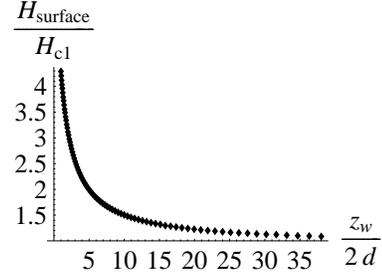}
\caption{Magnetic field at the surface at current approaching
$I_{c1}$ as function of $z_w$, for $x_w=2d$, $d = \lambda/2$.
\label{fg:Hratiozw}}
\end{center}
\end{figure}

Next we consider the dependence on $z_w$ for $z_w \ge x_w = 2 d$
which can be experimentally relevant. The limit $z_w \to \infty$ can
be treated analytically since the external field $\vec{H}_0$ is
uniform at all $x>(-d+\frac{x_w}{2})$. In this limit the maximal
magnetic field at the surface before vortices penetrate is
$H_{surface} (I_{c1}) \to H_{c1}$. For finite $z_w$ we calculated
$H_{surface} (I_{c1})$ numerically with the result plotted in
Fig.~(\ref{fg:Hratiozw}). Accordingly $z_w$ should not be too large
in order to obtain the effect discussed here including the
enhancement of the surface field in the vortex free state for a
disordered surface.
\section{Conclusions}
\label{se:Conclusions} In this work we studied solutions of London
theory in a geometry where an external mesoscopic current flows
parallel to a surface of a SC. Only above a threshold current
$I_{c0}$ there exist solutions with curved flux lines entering and
leaving the SC at the surface. At a larger threshold current,
$I_{c1}$, these solutions become energetically favorable, however an
energy barrier separates them from a vortex free solution. At a
third threshold current, $I_{S}$, this barrier disappears. To
determine the current at which vortices actually penetrate the
sample one has to account for the degree of disordered of the
surface. For strong surface disorder the vortex can penetrate at
$I=I_{c1}$ despite  the presence of the barrier, due to large local
magnetic fields produced at impurity sites allowing for nucleation
of vortices. On the other hand for clean surface the entrance of
vortices occurs at $I=I_S$.

By calculating those currents using a numerical solution of the
problem we conclude that for strong surface disorder the present
geometry allows to achieve locally larger magnetic fields in the
vortex-free state, as compared to the case of homogeneous magnetic
field, provided that the wire thickness is of O($\lambda$). This can
be potentially relevant for experiments in high-temperature
superconductors which typically have extremely low values of
$H_{c1}$. We argued that the effect of enhancement of the magnetic
field in the vortex free (Meissner) state becomes more pronounced in
strongly anisotropic superconductors, which is particularly relevant
for layered high-temperature superconductors.

We would like to thank Jordan Baglo, Walter Hardy and Cedric Lin for
stimulating discussions. This work was supported by NSERC (ES $\&$
IA) and CIfAR (IA).

\appendix
\section{Mixed terms in free energy}
\label{se:F0} We shall prove the vanishing of crossed term in the
energy Eq.~(\ref{eq:F}) between $\vec{H}_0$ and $\vec{H}_{s}$ [see
Eq.~(\ref{eq:H})],\begin{widetext}
\begin{eqnarray}
\label{eq:mixed} F_{(H_0 H_s)} = \mu_0 \int d^3 r \bigl( \vec{H}_0
\cdot \vec{H}_s +\theta(x)\lambda^2 (\vec{\nabla} \times \vec{H}_0)
\cdot (\vec{\nabla} \times H_s) \bigr)=0.
\end{eqnarray}
From Eq.~(\ref{eq:H}) we have $\vec{H}_0=\vec{H}_0'+\vec{H}_{s0}$
where $\vec{H}_0'= \theta(-x)(\vec{H}_{I,d}+\vec{H}_{-I,-d}) =
\vec{\nabla} \times \vec{A}_1$ and
\begin{equation}
\vec{A}_1 =\frac{I \hat{y} }{4 \pi } \log
\frac{(x+d)^2+z^2}{(x-d)^2+z^2} , ~~x<0.
\end{equation}
Correspondingly, we have $F_{(H_0 H_s)} =F_{(H_0' H_s)} +F_{(H_{s0}
H_s)} $. Consider the term $F_{(H_0' H_s)}=\mu_0  \int_{x<0} d^3 r
\vec{H}_0' \cdot \vec{H}_s$. We will use the vector identity
$(\vec{\nabla} \times \vec{A}) \cdot \vec{B} = \vec{A} \cdot
(\vec{\nabla} \times \vec{B})+\vec{\nabla} \cdot (\vec{A} \times
\vec{B})$, with $\vec{A}=\vec{A}_1$, $\vec{B}=\vec{H}_s$, and the
fact that $\vec{\nabla} \times \vec{H}_s=0$. Then the volume
integral can be transformed to an integral on the surface $x=0^-$.
However this integral vanishes because $\vec{A}_1 (0^-,y,z)=0$,
hence $F_{(H_0' H_s)} =0$. Now let us consider the term $F_{(H_{s0}
H_s)}$ and define $\vec{H}_s = \vec{\nabla} \times \vec{A}_s$. For
the integral in the region $x<0$ we use the above vector identity
with $\vec{A}=\vec{A}_s$, $\vec{B}=\vec{H}_{s0}$, and for $x>0$ we
use the vector identity with $\vec{A}=\vec{H}_{s0}$,
$\vec{B}=\vec{\nabla} \times \vec{H}_{s}$. Taking into account that
$\vec{H}_s$ and $\vec{H}_{s0}$ satisfy the homogeneous equations we
obtain
\begin{eqnarray}
\label{eq:xxx} \int_{x<0} d^3 r  (\vec{\nabla} \times \vec{A}_s)
\cdot \vec{H}_{s0}=\int dS \bigl(\vec{A}^-_s \times \vec{H}^-_{s0}
\bigr)_x, \nonumber \\ \int_{x>0} d^3 r  \bigr( \vec{H}_{s0} \cdot
\vec{H}_s+\lambda^2 (\vec{\nabla} \times \vec{H}_{s0}) \cdot
(\vec{\nabla} \times H_s) \bigl)=-\lambda^2\int dS
\bigl(\vec{H}^+_{s0} \times (\vec{\nabla} \times \vec{H}^+_{s})
\bigr)_x.
\end{eqnarray}
Here $\int dS = \int_{-\infty}^\infty dy \int_{-\infty}^\infty dz$,
and $\vec{H}^\pm = \vec{H}(x=0^\pm,y,z)$. For $x>0$ we can use
$\vec{A}_s = -\lambda^2 \vec{\nabla} \times \vec{H}_s$, which
follows from London equation for $\vec{H}_s$. Next we use the fact
that, by construction, $\vec{H}_s^- = \vec{H}_s^+ + \vec{H}_v^+$.
This allows us to express $\vec{A}_s^-=-\lambda^2 \vec{\nabla}
\times (\vec{H}_v^++ \vec{H}_s^+)$, and combine the two terms of
Eq.~(\ref{eq:xxx}) as
\begin{equation}
\label{eq:simple?} F_{(H_{s0} H_s)} = -\mu_0 \lambda^2 \int dS
\left[ (\vec{\nabla} \times \vec{H}^+_{s}) \times (\vec{H}_{s0}^-
-\vec{H}_{s0}^+ )+(\vec{\nabla} \times \vec{H}_{v})^+  \times
\vec{H}_{s0}^- \right]_x.
\end{equation}
Now we use explicit forms of these factors: $(\vec{H}_{s0}^-
-\vec{H}_{s0}^+ ) =-\frac{I d \hat{z}}{\pi (d^2+z^2)}$,
$(\vec{H}_{s0}^-)_y=0$,
\begin{eqnarray}
(\vec{\nabla} \times \vec{H}^+_{s})_y&=&-\frac{\phi_0}{2 \mu_0
\lambda^2} \int_\gamma d\vec{r}`_x \int \frac{d ^2 k_2}{(2 \pi)^2}
e^{- i \vec{k}_2 \cdot \vec{r}`+i(k_y y+k_z z)-\tau(k_2)
|\vec{r}`_x|} \frac{-i k_z(\tau(k_2) - k_2)}{k_2 \tau(k_2)}
,\nonumber \\
(\vec{\nabla} \times \vec{H}^+_{v})_y&=&\frac{\phi_0}{2 \mu_0
\lambda^2} \int_\gamma d\vec{r}`_x \int \frac{d ^2 k_2}{(2 \pi)^2}
e^{- i \vec{k}_2 \cdot \vec{r}`+i(k_y y+k_z z)-\tau(k_2)
|\vec{r}`_x|} \frac{i k_z}{\tau(k_2)}
\left(1-\frac{\tau^2(k_2)}{k_z^2}
\right),\nonumber \\
(\vec{H}_{s0}^-)_z&=&-I  \int \frac{d ^2 k_2}{(2 \pi)^2} e^{ i
\vec{k}_2 \cdot \vec{r}-|k_z|d} \frac{k_2 2 \pi \delta(k_y)
}{\tau(k_2)+k_2}.
\end{eqnarray}
Plugging these expressions in Eq.~(\ref{eq:simple?}), one can
readily obtain $F_{(H_{s0} H_s)}=0$ (without performing any
integration), completing the proof for $F_{(H_0 H_s)}
=0$.\end{widetext}


\begin{thebibliography}{99}
\bibitem{Bean64} C.P.~Bean and J.~D.~Livingston, Phys. Rev. Lett.~\textbf{12}, 14 (1964).
\bibitem{DeGennes66}P. G. De Gennes, Superconductivity of Metals and Alloys (Benjamin; 1966).
\bibitem{Joseph64} A.~S.~Joseph and W.~J.~Tomasch, Phys. Rev. Lett.~\textbf{12}, 219 (1964).
\bibitem{Blois64} R.~W.~De~Blois and W.~De~Sorbo, Phys. Rev. Lett.~\textbf{12}, 499 (1964).
\bibitem{Brandt80} E.~H.~Brandt, J. Low Temp. Phys. \textbf{ 42}, 557 (1981).
\bibitem{Sudbo91} A.~Sudb{\o} and E.~H.~Brandt, Phys. Rev. B \textbf{43}, 10482 (1991).
\bibitem{Yeshurun91} M.~Konczykowski, L.~I.~Burlachkov, Y.~Yeshurun, and F.~Holtzberg, Phys. Rev. B \textbf{43}, 13707 (1991).
\bibitem{Gregory01} J. K. Gregory \emph{et. al.}, Phys. Rev. B \textbf{64}, 134517 (2001).
\bibitem{Iniotakis08} C.~Iniotakis, T.~Dahm, and N.~Schopohl, Phys. Rev. Lett.~\textbf{100}, 037002 (2008).
\bibitem{Zeldov94} E.~Zeldov \emph{et. al.}, Phys. Rev. Lett.~\textbf{73}, 1428 (1994).
\bibitem{Brandt99} E.~H.~Brandt, Phys. Rev. B \textbf{60}, 11939 (1999).
\bibitem{Morozov97} N.~Morozov \emph{et. al.}, Physica C \textbf{291}, 113 (1997).
\bibitem{Altshuler95} E.~Altshuler and R.~Mulet, Journal of Superconductivity \textbf{8}, 779 (1995).
\bibitem{Shehata84} L.~N.~Shehata and A.~G.~Saif, J. Low Temp. Phys. \textbf{56}, 113 (1984).
\bibitem{Liang94} R.~Liang \emph{et. al.}, Phys. Rev. B \textbf{50}, 4212 (1994).
\bibitem{London50} F.~London, \emph{Superfluids} (Wiley, New York,
1950), Vol. I, Sec. VIII.
\bibitem{Bok68} J.~Bok and J.~Klein, Phys. Rev. Lett.~\textbf{20}, 660 (1968).
\bibitem{Brown68} J.~B.~Brown and T.~D.~Morris, \emph{Proc. 11th Int. Conf. Low. Temp.
Phys}., Vol. 2, 768 (St. Andrews, 1968).
\bibitem{Morris71} T.~D.~Morris and J.~B.~Brown, Physica (Amsterdam) \textbf{55}, 760
(1971).
\bibitem{Lipavsky04} P. Lipavsk\'{y} \emph{et. al.}, Phys. Rev. B \textbf{70}, 104518 (2004).
\bibitem{Blatter93} G.~Blatter and V.~Geshkenbein, Phys. Rev. B \textbf{47}, 2725 (1993),
\emph{ibid}. E.~H.~Brandt \textbf{48}, 6699 (1993).
\bibitem{Sudbo93} A.~Sudb{\o}, E.~H.~Brandt and D.~A.~Huse, Phys. Rev. Lett. \textbf{71}, 1451 (1993).
\end{thebibliography}
\end{document}